# Enhancing short-term traffic prediction by integrating trends and fluctuations with attention mechanism

Adway Das[a], Agnimitra Sengupta[b]* and S. Ilgin Guler[c]

[a,c] *Department of Civil and Environmental Engineering, Penn State University, University Park, United States;* [b]*Traffic Operations, HNTB Corporation, Bartow, United States*

*senguptaagni@gmail.com; asengupta@hntb.com

# Abstract

Traffic flow prediction is a critical component of intelligent transportation systems, yet accurately forecasting traffic remains challenging due to the interaction between long-term trends and short-term fluctuations. Standard deep learning models often struggle with these challenges because their architecture inherently smooths over fine-grained fluctuations while focusing on general trends. This limitation arises from low pass filtering effects, gate biases favoring stability, and memory update mechanisms that prioritize long-term information retention. To address these shortcomings, this study introduces a hybrid deep learning framework that integrates both long-term trend and short-term fluctuation using two input features processed in parallel designed to capture complementary aspects of traffic flow dynamics. Further, our approach leverages attention mechanisms, specifically Bahdanau attention, to selectively focus on critical time steps within traffic data, enhancing the model's ability to predict congestion and other transient phenomena. Experimental results demonstrate that features learned from both branches are complementary, significantly improving the goodness-of-fit statistics across multiple prediction horizons compared to a baseline model. Notably, the attention mechanism enhances short-term forecast accuracy by directly targeting immediate fluctuations, though challenges remain in fully integrating long-term trends. This framework can contribute to more effective congestion mitigation and urban mobility planning by advancing the robustness and precision of traffic prediction models

Keywords: traffic prediction, hybrid modeling, complementary feature, attention mechanism

**Introduction**

The effectiveness of traffic management systems relies on the accuracy of short/medium-term traffic flow predictions, which are critical for real-time traffic control and congestion mitigation strategies. These forecasts, typically spanning time horizons of five minutes to an hour, form the foundation for optimizing traffic operations and ensuring network reliability. Short-term forecasts enable the deployment of data-driven traffic control measures that mitigate congestion, enhance safety, and reduce environmental impacts. For instance, ramp metering systems rely on these forecasts to dynamically regulate vehicle entry onto freeways by adjusting signal timings at on-ramps, thereby improving merging efficiency and alleviating congestion on mainline traffic streams (1,2). Likewise, precise predictions are integral to detour management, enabling efficient rerouting of traffic away from incidents, construction zones, or high-demand corridors, thereby minimizing delays and maintaining network resilience. Adaptive traffic signal control systems further underscore the importance of short-term predictions, as they leverage real-time traffic conditions to dynamically adjust signal timings. These systems aim to minimize stop-and-go driving, optimize intersection throughput, and enhance fuel efficiency while reducing emissions (3,4).

Achieving high accuracy in traffic flow prediction is inherently challenging due to the dynamic and complex nature of traffic systems. Traffic flow is influenced by factors, such as road incidents, human driving behavior, and evolving travel demand patterns. Furthermore, the transition between free-flow and congested conditions introduces additional complexity to forecasting models (3, 5–7). In free-flow conditions, traffic moves predictably with minimal interactions among vehicles, whereas congested conditions are characterized by non-linear dynamics, including increased vehicle interactions, stop-and-go waves, and abrupt changes in speed and headway. These varying traffic regimes demand advanced modeling techniques capable of capturing the intricate relationships between traffic states and adapting to the highly variable nature of real-world traffic.

Short-term traffic prediction has traditionally relied on a variety of modeling approaches, including classical statistical models and modern machine learning (ML) techniques. Statistical parametric methods, such as historical average algorithms, smoothing techniques (8–10), and autoregressive integrated moving average (ARIMA) models, have been widely used to capture temporal fluctuations in traffic flow (11,12). However, these methods often struggle with real-world traffic complexities due to their rigid assumptions about data relationships. To overcome these limitations, non-parametric approaches, such as *k*-nearest neighbors, support vector machines, and Bayesian networks, have been introduced. These methods offer greater flexibility by avoiding predefined functional forms, enabling improved adaptability and performance across diverse and non-linear traffic datasets.

Deep learning (DL) techniques have significantly improved traffic prediction accuracy by modeling complex non-linear relationships (2). Recurrent neural networks (RNNs) and their variants, such as long short-term memory (LSTM) networks, are particularly effective in capturing temporal correlations in sequential data. These models have been widely applied to traffic forecasting tasks, including travel time and speed prediction, even under challenging conditions (13–16). However, conventional deep learning models often struggle with the non-Euclidean structures of traffic networks, limiting their ability to capture spatiotemporal dependencies. While models like BiLSTM (17), DELA (18), and ConvLSTM (19,20) excel in processing high-dimensional data, their performance in modeling spatiotemporal relationships on non-Euclidean networks remains suboptimal. Recent advancements in graph neural networks (GNNs) address this limitation by effectively modeling non-Euclidean structures, offering a robust framework for capturing complex spatiotemporal correlations and improving multi-step traffic flow prediction accuracy (21,22).

Hybrid models integrating spatial and temporal dependencies have significantly enhanced traffic flow prediction accuracy. Convolutional Neural Networks (CNNs) effectively capture spatial patterns, while Long Short-Term Memory (LSTM) networks excel at modeling temporal sequences. For instance, Sun et al. demonstrated that incorporating meteorological data into a CNN-LSTM model reduced mean prediction errors to below 5% (23). Similarly, Du et al.'s hybrid model combining spatial and temporal features outperformed traditional methods. However, many existing approaches primarily focus on spatio-temporal analysis, which may not fully address the unique characteristics of standalone or sparsely located vehicle detectors. Training individual detector stations independently using robust, generalized frameworks can better account for localized traffic flow variations (24). Recent studies, such as Yao et al., highlight the effectiveness of integrating attention mechanisms with local convolutional networks to capture long-term patterns and temporal changes, further enhancing prediction accuracy and reliability (25).

The shift toward more adaptable modeling approaches has improved the handling of diverse traffic scenarios, enhancing traffic forecasting and management. Duan et al. demonstrated that integrating reinforcement learning with hybrid models improves prediction accuracy and reduces execution time, even in complex deep networks (26). This approach overcomes limitations of traditional spatio-temporal models by enabling independent training for each detector station, better accounting for localized traffic patterns and anomalies. Similarly, Vijayalakshmi et al. emphasized the benefits of attention-based CNN-LSTM models that incorporate short-term flow characteristics and dependencies between speed and flow, achieving significant improvements in prediction precision (27). Emerging methods, such as fuzzy CNN (F-CNN) and transformer-based models, have further advanced spatiotemporal modeling. The F-CNN model by Duan et al. employs fuzzy inference to manage uncertain traffic incident data effectively, demonstrating superior performance on real-world datasets (26). Transformer-based models have also shown promise by accurately capturing temporal dependencies and avoiding the error accumulation typical of RNNs (28,29). These advancements underscore the importance of hybrid and dynamic modeling techniques in addressing real-world traffic complexities. Regardless, these models cannot accurately predict short-term fluctuations in traffic.

This paper introduces a novel hybrid deep learning framework for short-term traffic flow forecasting that addresses the limitations of standard LSTM models in capturing fine-grained short-term fluctuations. The primary contributions of this study are as follows:

- introducing a hybrid deep learning model that captures complementary aspects of traffic dynamics by combining long-term trend with short-term fluctuation
- conducting a systematic analysis to evaluate how attention mechanisms can enhance the model's ability to predict transient phenomena such as congestion

The structure of this paper is organized as follows: We first provide background information on the deep learning models and attention mechanisms employed in this study. Next, we describe our modeling approaches, and the data used. We then compare the performance of our proposed models with a baseline model across various prediction tasks. Finally, we present our concluding remarks based on the results.

**Introduction**

In this section, we provide an overview of the DL forecasting model i.e., long short-term memory (LSTM) and the Bahdanau attention mechanism used in this paper. Detailed descriptions of the complementary feature learning module employed in our study are provided in later sections.

*Long short-term memory*

Recurrent Neural Networks (RNNs) are specifically designed to handle sequential data by incorporating a feedback mechanism that allows information from previous time steps to influence current predictions. This recurrent structure enables RNNs to capture temporal dependencies, making them well-suited for modeling time series data, such as traffic patterns. However, standard RNNs struggle with learning long-term dependencies due to the vanishing gradient problem (30), which limits their ability to retain information over extended sequences. To address this issue, Long Short-Term Memory (LSTM) networks were introduced (31), incorporating memory cells and gating mechanisms that regulate information flow, allowing for the effective learning of long-range dependencies.

The goal is to use the input, $x_t$, (traffic measurements) and the hidden state, $h_{t-1}$, determined from the previous time step to update the cell, $C_t$, at time $t$. The input, $x_t$, and hidden state from $t-1$, $h_{t-1}$, are used to determine a candidate memory cell, $\tilde{C}_t$. Then, cell $C_t$ is updated using a forget gate, $f_t$, that selectively filters information from the past, $C_{t-1}$, and an input gate, $i_t$, that controls the amount of information from the candidate memory cell, $\tilde{C}_t$, that should be incorporated into the current cell state. Lastly, the update of the hidden state, $h_t$, is controlled by an output gate, $o_t$. The computations are represented as follows (refer to Figure 1):

$$\tilde{C}_t = \tanh(W_c[h_{t-1}, x_t] + b_c) \tag{1}$$

$$C_t = f_t \cdot C_{t-1} + i_t \cdot \tilde{C}_t$$

$$h_t = o_t \cdot \tanh(C_t)$$

The activations of the forget gate, $f_t$, input gate, $i_t$, and output gate, $o_t$ are computed as shown below:

$$f_t = \sigma(W_f[h_{t-1}, x_t] + b_f) \tag{2}$$

$$i_t = \sigma(W_i[h_{t-1}, x_t] + b_i)$$

$$o_t = \sigma(W_o[h_{t-1}, x_t] + b_o)$$

In these equations, $\sigma(\cdot)$ and $\tanh(\cdot)$ denote non-linear activation functions. The weight matrices $W_f, W_i, W_o$, and $W_c$ correspond to the forget gate, input gate, output gate, and candidate memory cell, whereas $b_f, b_i, b_o$, and $b_c$ represent the respective bias terms associated with each gate.

*Bahdanau Attention Mechanism*

The attention mechanism has transformed sequence modeling using NNs, enabling models to assign dynamic importance to input elements at each time step. This capability enhances the modeling of long-term dependencies and addresses the limitations of fixed-size context vectors in traditional sequence-to-sequence architectures. Attention mechanisms have been widely adopted in applications such as machine translation, speech recognition, and time series forecasting due to their ability to improve performance across diverse tasks (32-35)

The Bahdanau attention mechanism (36), also referred to as additive attention, enhances sequence-to-sequence models by enabling the decoder to focus selectively on relevant portions of the input sequence. By aligning the decoder's current state with encoder states dynamically, this method effectively captures dependencies in longer sequences, overcoming challenges faced by fixed-context representations.

In this approach, the encoder processes the input sequence $x = (x_1, x_2, \cdots x_T)$ to generate a sequence of hidden states $h = (h_1, h_2, \cdots, h_T)$. At each decoding step $t$, an alignment score $e_{ti}$ is computed between the $i$-th encoder hidden state $h_i$ and the decoder's previous hidden state $s_{t-1}$, using a feed-forward NN, as follows:

$$e_{ti} = v^T \tanh(W_a h_i + U_a s_{t-1}) \tag{3}$$

where $W_a$ and $U_a$ are the weight matrices applied to the encoder and decoder hidden states respectively, and $v^T$ is a learnable vector used to compute the scalar score. Here, $i$ indexes the encoder hidden states, while $t$ corresponds to the current output step of the decoder. This formulation allows to determine how each encoder state aligns with the output being generated at step $t$, enabling the model to focus on the most relevant parts of the input sequence dynamically.

The alignment scores are then normalized into attention weights $\alpha_{ti}$ using the softmax function.

$$\alpha_{ti} = \text{softmax}(e_{ti}) \equiv \frac{\exp(e_{ti})}{\sum_{k=1}^{T_x} \exp(e_{tk})} \tag{4}$$

These attention weights ($\alpha_{ti}$) represent the relative importance of each encoder hidden state $h_i$ for predicting the decoder output at time step $t$. Using these weights, the context vector $c_t$, which captures the most relevant parts of the input sequence, is computed as:

$$c_t = \sum_{i=1}^{T_x} \alpha_{ti} h_i \tag{5}$$

where, $T_x$ is the length of the input sequence. Once the context vector $c_t$ is obtained, it is concatenated with the decoder's previous hidden state $s_{t-1}$, to form a combined vector: $[c_t; s_{t-1}]$. This combined vector is then passed through a dense layer with 'tanh' activation to compute the final attention vector, $a_t$:

$$a_t = \tanh(W_c[c_t; s_{t-1}]) \tag{6}$$

Depending on the specific model architecture, attention vectors can be processed differently to generate predictions. They are often input to dense or recurrent layers for generating predictions. In our models, as detailed later, the attention vectors are passed through dense layers, followed by a prediction layer, ensuring the effective use of context captured by the attention mechanism.

**Methodology**

In this study, we propose a hybrid deep learning model that leverages complementary features to enhance traffic flow prediction accuracy. In traffic flow prediction, integrating features that represent distinct temporal dynamics—such as long-term trends and short-term variations—enables the model to capture a more comprehensive view of traffic behavior. Long-term trends reflect stable patterns like daily or seasonal variations, while short-term fluctuations capture rapid changes caused by events such as differential vehicle arrival rate, congestion or signal timing. By dynamically weighing the contributions of each component, this approach has been demonstrated to improve the model's ability to generalize across diverse traffic conditions and handle non-linearities and temporal dependencies effectively. For example, one study combined features derived from traffic flow data with latent state transition probabilities to effectively capture short-term variations, achieving superior performance compared to other methods (37). Another study employed wavelet decomposition to separate long-term trends from detailed components, processing each through separate neural network branches and subsequently merging the features for improved prediction accuracy (38).

Mathematically, the combined feature set $X_{CFE}$ can be formulated as:

$$X_{CFE} = \alpha X_{\text{long-term}} + \beta X_{\text{short-term}} \tag{7}$$

where, $X_{\text{long-term}}$ and $X_{\text{short-term}}$ represent the long-term flow patterns and short-term fluctuations respectively, and $\alpha$ and $\beta$ are weights that balance their contributions. These weights can be learned dynamically during model training to optimize predictive performance.

In this study, *traffic flow* serves as the long-term feature, while *vehicle count fluctuations over time* (i.e., changes in vehicle arrival at each step) represent the short-term feature. The complementary features are processed through distinct model components to extract meaningful patterns, which are then fused for final prediction. By focusing on these complementary aspects, the proposed approach enhances the model's ability to predict both stable trends and transient phenomena, such as sudden congestion spikes.

*Proposed Model Architecture*

Model architecture for the proposed model is shown in Figure 2. As can be seen, two distinct feature sets—traffic flow data and flow fluctuation data—are processed (encoded) using two separate LSTM networks. Each LSTM network is comprised of two layers: the first layer has 25 units, and the second has 15 units, both utilizing ReLU activation functions. These layers extract temporal patterns and provide deeper abstractions of the input sequences while preserving the unique characteristics of each data source.

To enhance the model's ability to capture relevant temporal information, an attention mechanism inspired by the Bahdanau framework is incorporated. This mechanism computes alignment scores between the encoder's (LSTM in Figure 2) hidden states and the decoder's (Dense layer in Figure 2) previous hidden state, enabling the model to dynamically focus on the most relevant temporal patterns in the input data.

The outputs of the two LSTM networks are passed through independent attention layers, which compute attention vectors specific to each feature set, as shown below.

$$a_t^X = \tanh(W_c^X[c_t^X; s_{t-1}^X]), \quad a_t^{\Delta X} = tanh\big(W_c^{\Delta X}[c_t^{\Delta X}; s_{t-1}^{\Delta X}]\big) \tag{8}$$

where $a_t^X$ and $a_t^{\Delta X}$ are the attention vectors corresponding to traffic flow and flow fluctuation data, respectively. This design ensures that the model maintains and leverages the distinct temporal patterns present in each feature set.

These attention vectors are then passed through dense layers (the decoder) to derive high-level representations. Each feature set is processed through a dense layer containing 15 units with LeakyReLU activation, capturing non-linear relationships in the data. The resulting outputs are concatenated to form a unified representation, integrating complementary information from both feature sets. This combined representation undergoes further processing through an additional dense layer with 10 units and LeakyReLU activation. Finally, the output layer predicts traffic flow for the specified lead time steps, utilizing the integrated features derived from both traffic flow and flow fluctuation data.

## *Data Description*

The data for this study was obtained from the California Department of Transportation's Performance Measurement System (PeMS). Vehicle detector sensors (VDS) installed along freeway lanes and ramps recorded key traffic parameters, including flow, occupancy, and speed. The original 30-second interval data was processed into 5-minute aggregates for analysis. This study focused on one year of flow data collected from VDS 1114805 on California Interstate-5 NB in District 11, totaling 104,942 data points.

To prepare the dataset for supervised learning using LSTM-based networks, historical traffic data were converted into overlapping input-output pairs, where the past 10-time steps (lag) were used as input to predict the subsequent 5-time steps (lead). This transformation preserved temporal dependencies and enabled the model to learn traffic patterns effectively. The structured sequences were reshaped to match the input format required by LSTM networks, ensuring robust and efficient processing of time series data for accurate traffic flow forecasting. Additionally, fluctuation data, representing variations in traffic flow, was processed similarly and input as a secondary feature in our proposed model to efficiently model short-term variations in traffic. Figure 3 depicts the fluctuation data for a typical 48-hour period observed at the detector, highlighting the variability in traffic flow over time.

**Results and discussion**

To evaluate the proposed model's performance, we systematically analyze the impact of merging traffic flow and fluctuation data and the role of attention mechanisms. Several benchmarks are considered. A Regular LSTM Model, which processes only traffic flow data through a single LSTM network is developed and compared to a hybrid model without an attention mechanism – referred to as *Merged*. The merged model incorporates two branches: one for traffic flow data and another for flow fluctuation data, both processed through separate LSTM networks. The outputs from the two LSTM branches are concatenated and passed through dense layers with LeakyReLU activation to produce predictions. This comparison allows us to assess how integrating complementary features improves predictive performance over a single-feature model.

Next, we explore the contribution of attention mechanisms by integrating them into both the regular LSTM model (referred to as *LSTM+Attn*) and the merged model (referred to as *Merged+Attn*). In the regular LSTM Model with attention, a single attention mechanism is applied to the traffic flow data to capture dynamic temporal dependencies. In the merged model with attention, each branch is equipped with its own attention mechanism tailored to its respective feature set, enabling the model to selectively focus on relevant temporal patterns within each data type. To further analyze the impact of applying attention mechanisms to different feature sets, we evaluate two variants of the merged model. The first variant, *Merged+Attn*(flow), applies attention only to the traffic flow branch, while the second variant, *Merged+Attn*(flow, fluctuation), incorporates separate attention mechanisms for both the traffic flow and flow fluctuation branches.

The input to each model consists of 20-time steps, which are used to predict the next 5-time steps, corresponding to a 25-minute forecast. To assess prediction accuracy, we compare the predicted values with the actual flow measurements using two performance metrics: root mean squared error (RMSE) and mean absolute percentage error (MAPE), as defined below:

$$RMSE = \sqrt{\frac{1}{N}\sum_{i=1}^{N}(y_i - \widehat{y_i})^2} \tag{9}$$

$$MAPE = \frac{1}{N}\sum_{i=1}^{N}\left|\frac{y_i - \widehat{y_i}}{y_i}\right|^2 \tag{10}$$

where $y_i$ represents the true value of the flow observation $i$ (in units of veh/5min); $\widehat{y_i}$ is the predicted value of $y_i$, for all $i = 1,2,\cdots,N$. We selected RMSE as the primary evaluation metric due to its higher sensitivity to larger errors, larger errors, making it more sensitive to outliers compared to other metrics like MAE. This characteristic makes RMSE particularly suitable for our analysis, as it emphasizes more significant deviations in prediction accuracy.

The models were trained on 60% of the dataset and validated on 15%, with the remaining 25% held out for final evaluation. During the training process, the mean squared error loss function was minimized over 300 epochs, and the model with the lowest validation error was chosen to mitigate the risks of overfitting or underfitting. A grid search was performed with Adadelta (Zeiler, 2012) as the optimizer, and $\rho$ value of 0.95, and $\varepsilon$ of 1e-7, to optimize the learning rate and batch size, ensuring the best combination for model performance. The selected parameters are shown in Table 1.

*Performance comparison*

The experimental results highlight the impact of merging traffic flow and fluctuation features, as well as the contribution of attention mechanisms in improving predictive accuracy at various time horizons. Figures 4, 5, and 6 illustrate the models' performance. Figure 4 shows sample predictions for the 1st and 5th-time steps from each model, while Figures 5 and 6 compare RMSE and MAPE, respectively. We can observe a clear pattern across all models: prediction errors tend to increase as the forecast horizon extends, with substantially higher errors observed at the 5th time step compared to the 1st. This trend aligns with the inherent difficulty of longer-term predictions, where accumulated uncertainties lead to reduced forecasting accuracy.

First, the benefits of combining both the traffic flow data and flow fluctuation data are explored by comparing the results of the *LSTM* model compared to the *Merged* model. The results suggest that the *Merged* models demonstrate a superior ability to capture the true trends and finer fluctuations in the data compared to the regular *LSTM* model, regardless of whether equipped with attention mechanisms or not. This performance boost can be attributed to the inclusion of flow fluctuation data as an additional feature, which enables the models to better focus on short-term variations in traffic flow. While flow fluctuation data may appear as a linear transformation of flow data, its explicit inclusion as a separate input allows the models to extract complementary information. This enhances the models' ability to predict subtle variations, particularly at shorter prediction horizons where short-term dynamics dominate.

On longer time horizons, however, the dominant trends in the flow data become more influential. This shift reduces the relative impact of the fluctuation data, emphasizing the need for models to balance short- and long-term dynamics effectively.

Nevertheless, the merged models consistently outperform the standalone LSTM, showcasing the utility of leveraging distinct feature sets for improved prediction.

The impact of adding attention mechanisms to the models is less pronounced but still observable, e.g., see *Merged* vs. *Merged+Attn* or *LSTM* vs. *LSTM+Attn* models. For the standalone LSTM, attention provides minor improvements, particularly at longer horizons, where capturing relevant temporal dependencies becomes more challenging. Similarly, for the merged models, the inclusion of attention yields modest performance gains. This limited impact might stem from the fact that the merged models already benefit from the additional information provided by the flow fluctuation data, leaving limited scope for attention mechanisms to further enhance performance.

In summary, the results underscore the significant contribution of incorporating flow fluctuation data in improving predictive accuracy, particularly for short-term forecasts. While attention mechanisms offer more incremental benefits, the primary advantage lies in leveraging complementary features that effectively capture both short-term variations and long-term trends in traffic flow data, improving predictions especially in the long-term.

We also explore how attention mechanisms impact model performance during **peak periods**, characterized by congestion. Congested conditions, marked by non-linear dynamics such as stop-and-go waves, abrupt speed changes, and increased vehicle interactions, pose significant challenges for accurate traffic prediction. The experimental results reveal that the minor overall improvements in RMSE achieved by adding attention are largely attributed to its performance during these congested periods, as illustrated in Figure 7. For instance, while attention provides limited benefits for single-step predictions in the LSTM model, its contributions become increasingly significant across longer time horizons. Similarly, the merged models with attention also exhibit enhanced performance during congestion, with the advantages of attention becoming more pronounced as the prediction horizon extends. Notably, the models were trained to minimize MSE across all time steps and traffic conditions, rather than exclusively focusing on congestion. While this general optimization approach ensures balanced model performance, future work could explore modifying the loss function to differentially prioritize congestion and free-flow conditions, potentially leading to further improvements in congestion-specific predictions.

*Exploring the model-based benefits of merged models and attention mechanisms*

To substantiate the hypothesis that merging traffic flow and flow fluctuation data facilitates complementary feature learning, an in-depth analysis of the feature representations generated by the models was performed. Feature outputs were extracted from the dense layers of each branch prior to their concatenation. These high-dimensional feature representations were then visualized using t-SNE (t-distributed Stochastic Neighbor Embedding) to evaluate their separability. To distinguish features derived from flow data and flow fluctuation data, distinct color codes were employed. Minimal overlap between the feature sets would indicate that the models successfully captured distinct and complementary information from the two data sources. The t-SNE visualizations, as shown in Figure 8, revealed minimal overlap, demonstrating that the models effectively learned non-redundant features from the merged inputs. This result supports the hypothesis that incorporating flow fluctuation data enhances the models' ability to extract additional, unique information, which contributes to a more comprehensive representation of traffic dynamics and improved predictive performance.

Figures 9 and 10 presents color maps of normalized gradients across time for the three models, corresponding to different input vectors over a 24-hour period. The x-axis represents the feature index, which corresponds to the input historical flow data from previous time steps (with the most recent data on the right-hand side), while the y-axis represents different times of the day. This gradient analysis allows us to identify patterns that influence model performance, particularly for merged models, and to examine how attention mechanisms affect the temporal focus of the models.

For the regular LSTM model, gradients are notably higher during the increasing flow period around the 6-hour mark, coinciding with the rise in traffic toward peak periods. Outside these peak transition periods, gradients remain relatively low. Similarly, there seems to be some small increased gradients for the more recent data. Introducing attention to this model does not yield significant changes in gradient patterns, indicating limited impact of attention in a single-branch architecture.

By contrast, the hybrid (merged) models exhibit two distinct gradient maps, corresponding to the flow and flow fluctuation branches. For the regular merged model, gradients for the flow branch are concentrated on the most recent 10-time steps for most of the day, highlighting the model's focus on short-term temporal dependencies. This differs from the regular LSTM, where the focus is more diffuse. Interestingly, the sharp rise in traffic flow observed at the 6-hour mark is strongly reflected in the flow fluctuation branch, where gradients are high, indicating that the addition of fluctuation data enables the model to capture rapid transitions in traffic conditions.

The introduction of attention mechanisms further modifies the temporal focus. For the merged model with attention applied to the flow branch, gradients extend further back in time, allowing the model to capture long-term dependencies, which contributes to improved performance, as evidenced by lower RMSE values especially in the long term. Adding attention to the flow fluctuation branch enhances the model's ability to selectively focus on critical transition periods, such as congestion at 6 hours, leading to more pronounced gradient patterns during periods of high nonlinearity in traffic dynamics.

While these results highlight the strengths of the proposed hybrid models, they also reveal some limitations. For instance, the attention mechanisms are uniformly applied across all time steps, potentially neglecting variations in temporal importance specific to different times of the day. Additionally, the models are optimized for overall performance across all time steps, which may lead to suboptimal predictions during critical periods, such as peak hours. Future work could explore dynamic attention mechanisms that adapt based on temporal contexts and modified loss functions that prioritize high-accuracy predictions during congested periods. These enhancements could further improve the applicability of such models for real-world traffic forecasting tasks.

**Conclusions**

Accurately predicting traffic flow remains a complex task, primarily due to the inability of traditional deep learning models, such as LSTMs, to effectively capture short-term fluctuations alongside long-term trends. These models inherently smooth over fine-grained variations by prioritizing stability and long-term retention, which limits their ability to forecast rapid transitions and congestion periods. To overcome these limitations, this study introduced a hybrid deep learning framework that combines traffic flow data with its short-term fluctuations, employing attention mechanisms to selectively highlight critical time steps and transient patterns.

Our findings demonstrate that integrating flow and fluctuation features enables complementary feature learning, allowing the model to capture both gradual and abrupt traffic dynamics. The incorporation of attention mechanisms enhances this process by dynamically focusing on key time steps, enabling the model to adapt to complex conditions, such as congestion. Empirical evaluations reveal that the proposed framework significantly outperforms baseline LSTM models across multiple prediction horizons, particularly during periods of congestion, where traditional models often falter due to non-linear and abrupt changes in traffic conditions. Gradient analyses underscore how merging fluctuation data and utilizing attention mechanisms shift the model's temporal focus, facilitating a balanced representation of both short-term and long-term dependencies. Additionally, t-SNE visualizations support the complementary nature of features derived from flow and fluctuation data, highlighting their distinct contributions to improved prediction accuracy. These results affirm the value of incorporating short-term fluctuations as a separate input, despite being derived from flow data, as they provide unique insights that enhance overall model performance.

While the proposed framework represents a significant step forward, challenges remain in optimizing performance across diverse time horizons. A uniform loss function, as employed in this study, may not sufficiently prioritize critical periods such as peak traffic hours, leading to potential accuracy trade-offs. Future research will focus on developing temporally adaptive attention mechanisms and weighted loss functions to better balance the contributions of short-term and long-term data. Additionally, incorporating dynamic attention mechanisms that adjust focus based on real-time traffic patterns and contextual conditions holds promise for further enhancing the robustness and adaptability of the framework. By addressing the limitations of traditional models and advancing the integration of complementary features, this study contributes to the development of more precise and adaptable traffic prediction systems for improved congestion management, proactive traffic control, and sustainable urban mobility planning.

**Declaration of conflicting interests**

The author(s) declared no potential conflicts of interest concerning the research, authorship, and/or publication of this article.

**Tables**

Table 1: Hyperparameter optimization

| Model name | Batch size | Learning rate | Min val. loss |
|---|---|---|---|
| LSTM | 15 | 0.1 | 0.0384 |
| LSTM + Attn | 15 | 0.1 | 0.0386 |
| Merged | 15 | 0.1 | 0.0295 |
| Merged + Attn (Flow) | 15 | 0.1 | 0.0295 |
| Merged +Attn (Flow, Fluctuation) | 20 | 0.1 | 0.0294 |

# Figures

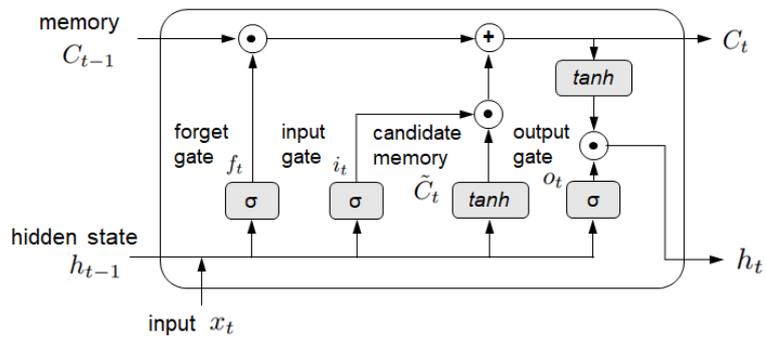

Figure 1. Schematic Diagram of an LSTM Module

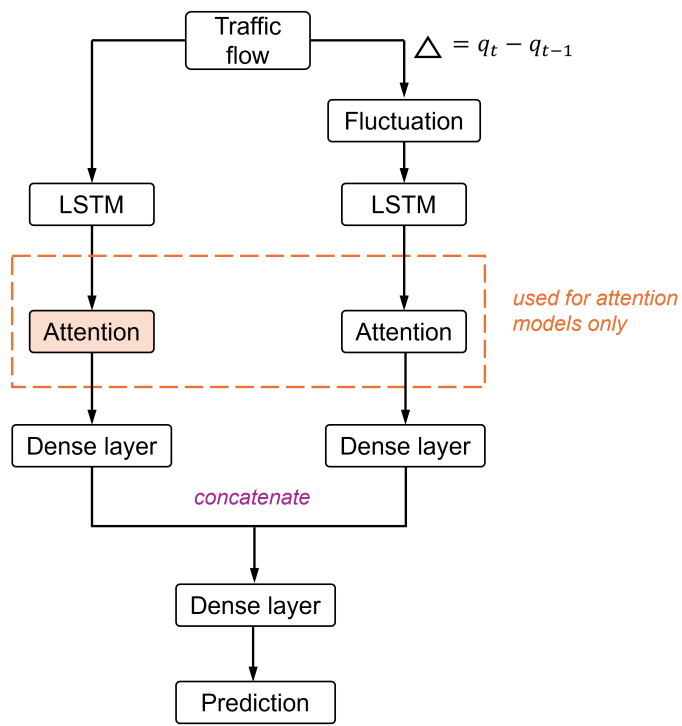

Figure 2. Model architecture of the proposed complementary learning model

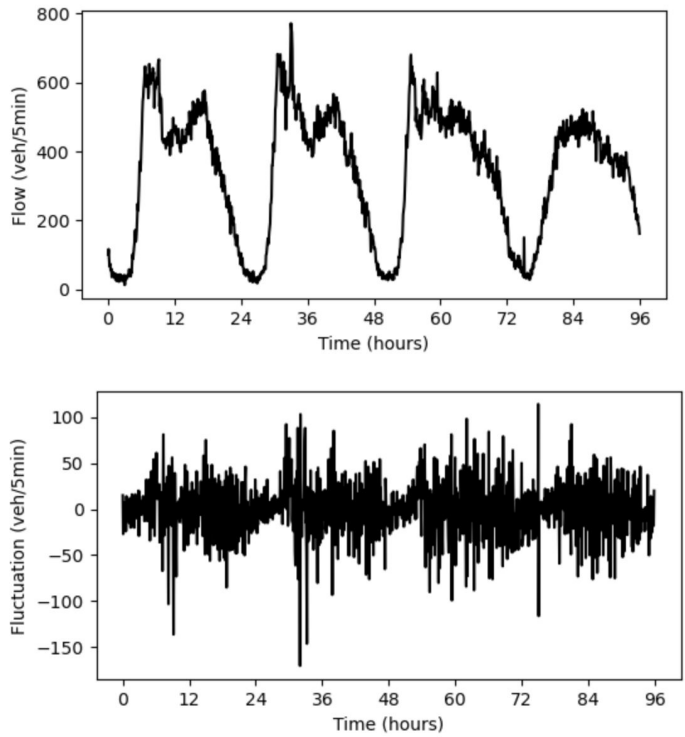

Figure 3. (a) Traffic flow patterns and (b) fluctuation for 48 Hours

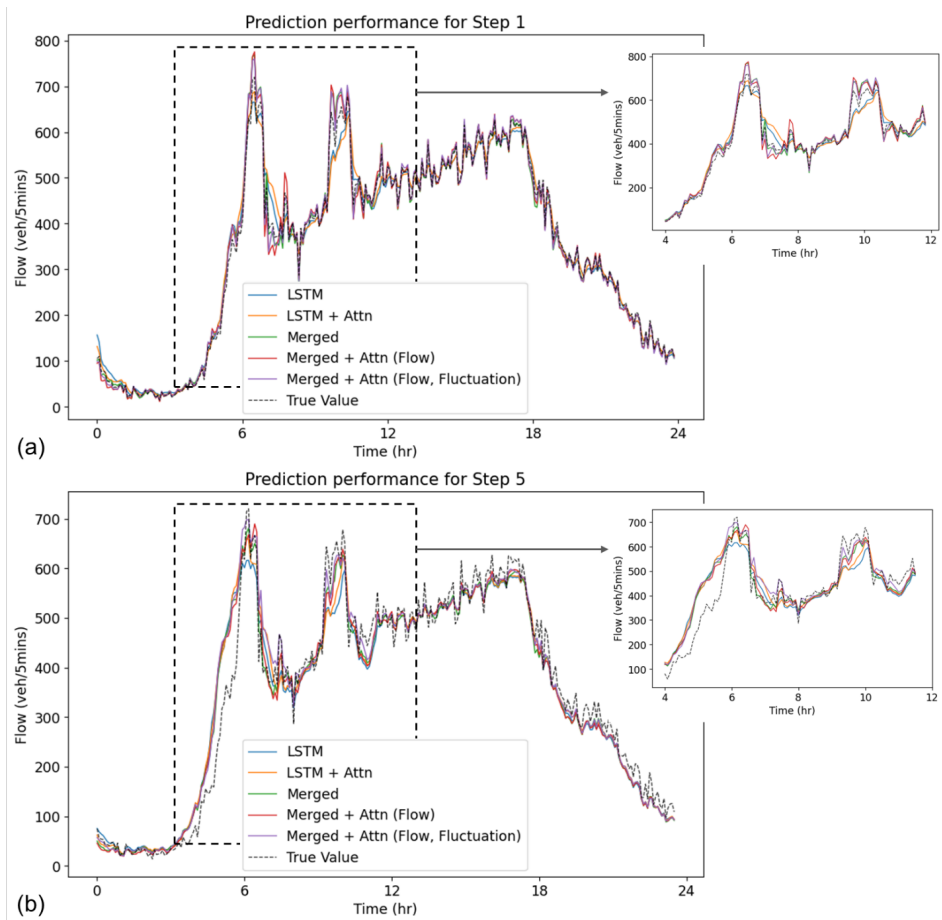

Figure 4: Typical comparison of flow prediction for the (a) 1st and (b) 5th time steps

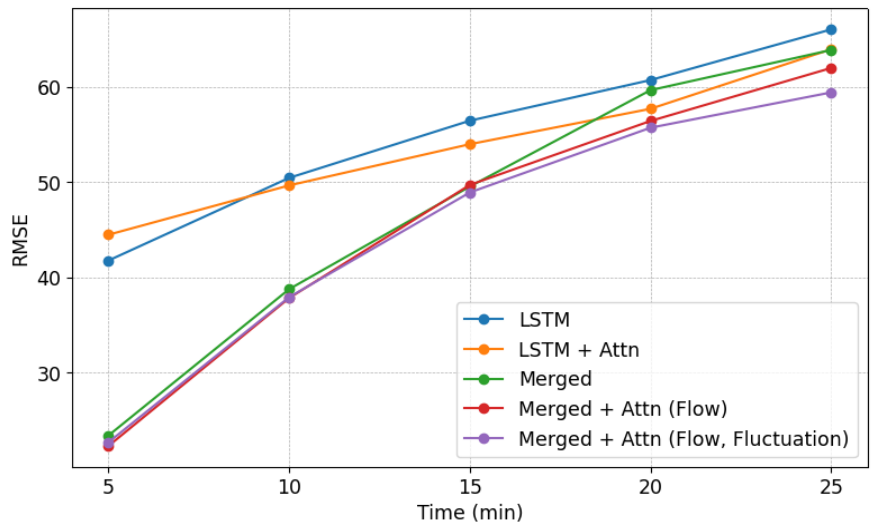

Figure 5. Overall RMSE across different forecast steps

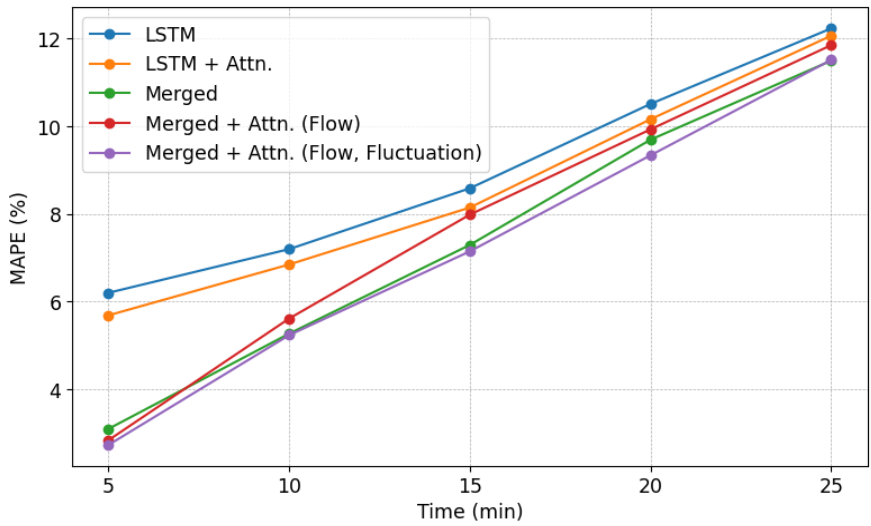

Figure 6. Overall MAPE across different forecast steps

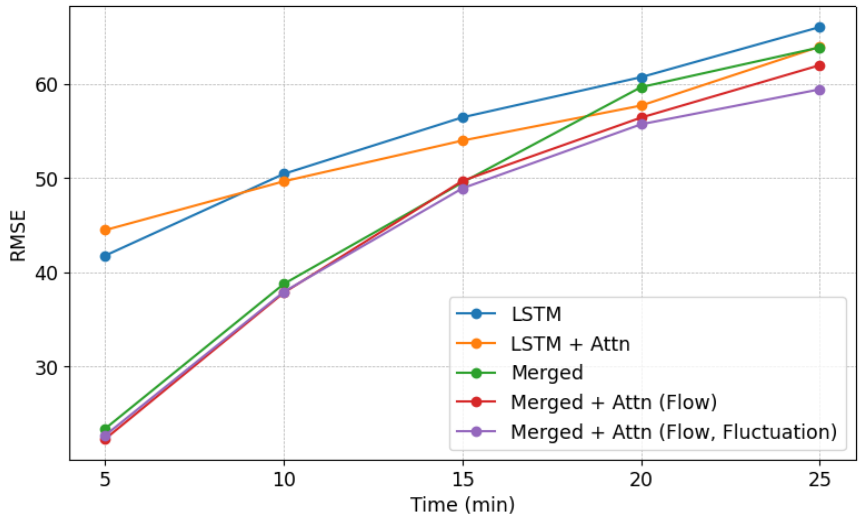

Figure 7. RMSE comparison for congested periods

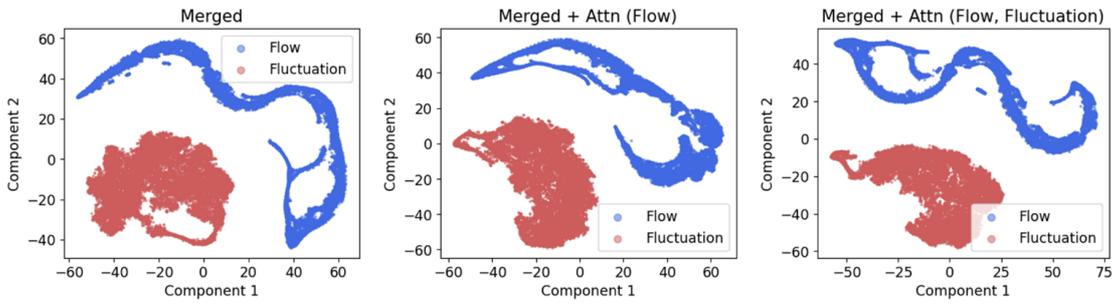

Figure 8. Latent space representations of features prior to concatenation using t-SNE

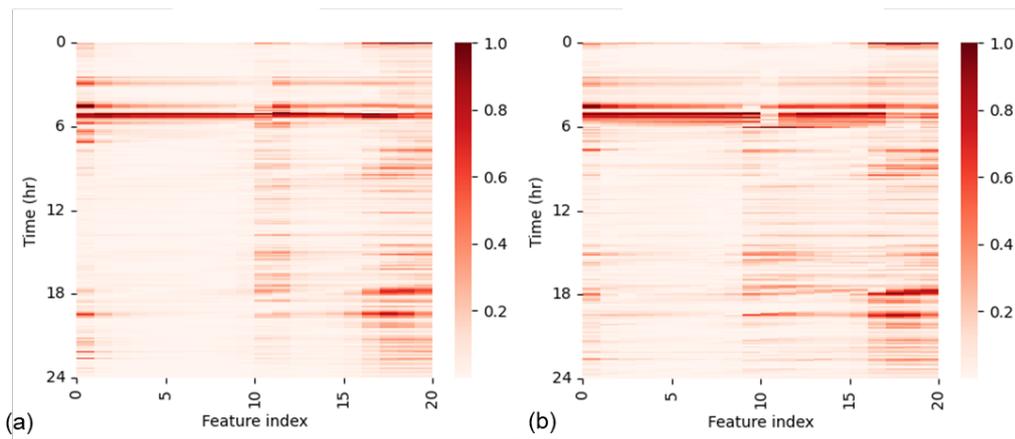

Figure 9. Gradient outputs for (a) regular LSTM and (b) LSTM with attention

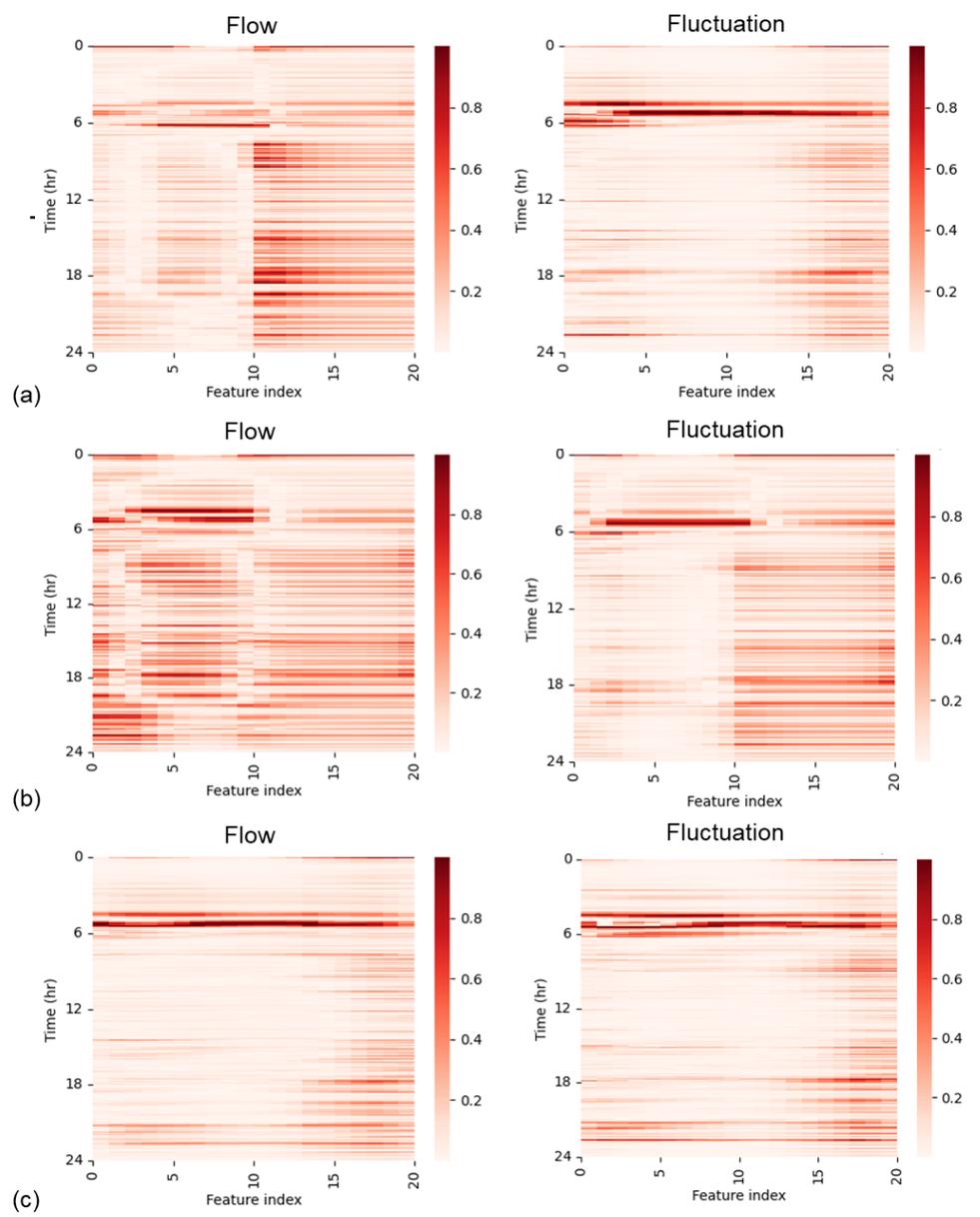

Figure 10. Gradient outputs for hybrid models (a) merged model; (b) merged model with attention on flow; (c) merged model with attention on flow and fluctuation

**Figure captions**

Figure 1. Schematic Diagram of an LSTM Module

Figure 2. Model architecture of the proposed complementary learning model

Figure 3. (a) Traffic flow patterns and (b) fluctuation for 48 Hours

Figure 4: Typical comparison of flow prediction for the (a) 1st and (b) 5th time steps

Figure 5. Overall RMSE across different forecast steps

Figure 6. Overall MAPE across different forecast steps

Figure 7. RMSE comparison for congested periods

Figure 8. Feature space visualizations of layer outputs prior to concatenation

Figure 9. Gradient outputs for (a) regular LSTM and (b) LSTM with attention

Figure 10. Gradient outputs for hybrid models (a) merged model; (b) merged model with attention on flow; (c) merged model with attention on flow and fluctuation